\documentclass[a4paper,11pt]{article}
\pdfoutput=1 

\usepackage{jheppub}

\usepackage[T1]{fontenc}

\usepackage{feynmp-auto}
\usepackage{subcaption}
\usepackage{tikz}
\usetikzlibrary{shapes,arrows}

\tikzstyle{block} = [rectangle, draw, text centered, minimum height=2em]
\tikzstyle{line} = [draw, -latex']

\renewcommand{\d}[2][]{\mathrm{d}^{#1}{#2}}
\newcommand{\del}[3][]{\frac{\partial^{#1}{#2}}{\partial{#3}^{#1}}}
\newcommand{\de}[3][]{\frac{\mathrm{d}^{#1}{#2}}{\mathrm{d}{#3}^{#1}}}

\allowdisplaybreaks

\title{Thermodynamics of 4D Dilatonic Black Holes and the Weak Gravity Conjecture}

\author[a]{Gregory J. Loges,}
\author[b]{Toshifumi Noumi}
\author[a]{and Gary Shiu}

\affiliation[a]{Department of Physics, University of Wisconsin-Madison, Madison, WI 53706, USA}
\affiliation[b]{Department of Physics, Kobe University, Kobe 657-8501, Japan}

\emailAdd{gloges@wisc.edu}
\emailAdd{tnoumi@phys.sci.kobe-u.ac.jp}
\emailAdd{shiu@physics.wisc.edu}

\preprint{KOBE-COSMO-19-15, MAD-TH-19-07}

\abstract{
Taking a thermodynamic perspective, we study the weak gravity conjecture in the context of 4D Einstein-Maxwell-dilaton theory. We find closed-form expressions for the corrected thermodynamic quantities in the presence of four-derivative terms in the action, and in particular the charge-to-mass ratio and entropy, for several families of solutions of special magnetic-to-electric charge ratio or dilaton coupling constant. Assuming that dyonic black holes themselves are the conjectured charged states, this places constraints on the Wilson coefficients of the theory which we show are satisfied under mild assumptions on the UV theory.
}

\begin{document}

\maketitle
\flushbottom

\section{Introduction}
\label{sec:Intro}
Classically, black holes have a well-defined temperature and thermodynamic entropy and are the source of several puzzles, such as the information paradox and the origin of their area-law entropy. String theory has offered explanations and explicit microscopic counting of states which give rise to the area-dependence for these solutions \cite{Strominger:1996sh}, a property which is seemingly mysterious when viewed from the perspective of classical Einstein gravity. Similarly, questions surrounding the infinite-density singularities of the classical theory are expected to be resolved by higher-energy physics, be it through stringy or, more generally, quantum gravity effects. One may systematically characterize these effects using the techniques of effective field theory, where a derivative expansion captures those features of the UV theory which are relevant for lower-energy processes. Of course, the challenge is to sum over the infinitely many higher derivative corrections in situations where this is necessary, such as in addressing the aforementioned cosmic singularities. String theory provides a UV complete framework in which such summation can at least in principle be carried out, as manifest e.g.~in the Regge behavior of string scattering amplitudes. 

Black holes also feature in the swampland program~\cite{Vafa:2005ui}, where they play a central role in thought experiments which provide a motivation for the weak gravity conjecture (WGC) \cite{ArkaniHamed:2006dz}. For a recent review of the swampland program, including the WGC, see \cite{Brennan:2017rbf,Palti:2019pca}. The WGC states that effective theories which have a UV completion including gravity must contain a state with charge-to-mass ratio greater than one, in appropriate units. It was quickly noted that black holes themselves may satisfy the conjecture if higher-derivative corrections increase the charge-to-mass ratio of extremal black hole states from their classical value of one \cite{Kats:2006xp}. For black holes at least, the WGC is equivalent to requiring that two identical extremal black holes repel one another,
\begin{equation}
    |F_\text{EM}|>|F_\text{grav}|+|F_\text{scalar}|\,,
\end{equation}
and hence have the decay to smaller black holes being kinematically allowed (in general the WGC and ``Repulsive Force Conjecture'' are not equivalent \cite{Heidenreich:2019zkl}, but such distinction is not relevant for our discussion). In 4D with Coulombic forces, this becomes
\begin{equation}\label{eq:WGCcharges}
    M^2 < 2M_\text{Pl}^2(Q_e^2 + Q_m^2 - Q_\phi^2) \,,
\end{equation}
where $M$ represents the black hole's mass and $Q_e$, $Q_m$ and $Q_\phi$ represent its electric, magnetic and scalar charges, respectively. The weak version of the WGC has been demonstrated, under some assumptions\footnote{See our comment in section~\ref{sec:Discussion}.}, in Einstein-Maxwell theory \cite{Hamada:2018dde,Charles:2019qqt,Jones:2019nev,Bellazzini:2019xts}.

It is natural to extend these arguments to Einstein-Maxwell-dilaton (EMd) theory, which arises as the low-energy theory of both KK reduction and string theory. With a massless dilaton the low-energy effective theory now includes more degrees of freedom than Einstein-Maxwell theory, and one cannot integrate out the dilaton to work in an EFT of only gravitons and photons. In this paper, we leverage the thermodynamic properties of black holes to derive bounds on the Wilson coefficients of the low-energy effective theory for several important EMd black hole solutions and show that the bounds are satisfied under generic assumptions on the UV theory.

This paper is organized as follows. In section~\ref{sec:BHreview} we review 4D EMd black holes solutions and their thermodynamics. In section~\ref{sec:ThermoOutline} we outline the technique used to extract higher-derivative corrections to all thermodynamic quantities. In section~\ref{sec:MassEntropyCorrections} we present the corrections to the charge-to-mass ratio and entropy which are relevant for the WGC. In section~\ref{sec:WGC} we show that the derived bounds are indeed satisfied under mild assumptions on the UV theory, and we conclude in section~\ref{sec:Discussion}.

\section{Review of Einstein-Maxwell-dilaton Black Hole Solutions}
\label{sec:BHreview}

We begin by reviewing static, dyonic black hole solutions of 4D EMd theory, for which we write the action as
\begin{equation}\label{eq:EMdAction}
    I = \int\d[4]{x}\,\sqrt{-g}\left[\frac{M_\text{Pl}^2}{2}R - \frac{M_\text{Pl}^2}{2}(\partial\phi)^2 - \frac{1}{4}e^{-2\lambda\phi}\big(F^2\big)\right] \,.
\end{equation}
Here and in what follows we use the compact notation $(\partial\phi)^2=\partial_\mu\phi\partial^\mu\phi$ and $(F^2)=F_{\mu\nu}F^{\mu\nu}$. Going forward we will set $M_\text{Pl}^2=1/8\pi G_\text{N}=1$. The exponential coupling constant, $\lambda$, may take on any real value, and it will be convenient to introduce the associated constant $h=\frac{2}{1+2\lambda^2}\in(0,2]$. Several special values for $\lambda$ are of note: $\lambda=0$ ($h=2$) gives Einstein-Maxwell theory with a decoupled dilaton; $\lambda^2=1/2$ ($h=1$) corresponds to the low-energy effective action of string theory; $\lambda^2=3/2$ ($h=1/2$) corresponds to the KK reduction of Einstein gravity from 5D to 4D, where the radion plays the role of the dilaton.

The action/equations of motion of \eqref{eq:EMdAction} enjoy two dualities:
\begin{equation}\label{eq:Dualities}
\begin{aligned}
    (\lambda,\phi) &\to (-\lambda,-\phi)\,,\\
    (F_{\mu\nu},\widetilde{F}_{\mu\nu},\phi)&\to(\widetilde{F}_{\mu\nu},-F_{\mu\nu},-\phi)\,,
\end{aligned}
\end{equation}
where $\widetilde{F}_{\mu\nu}\equiv \frac{1}{2}e^{-2\lambda\phi}\epsilon_{\mu\nu\rho\sigma}F^{\rho\sigma}$. Under the second duality, which we refer to as electromagnetic duality, electric and magnetic charges are interchanged as $(q_e,q_m)\to(q_m,-q_e)$.

A spherically-symmetric, static solution of the equations of motion is \cite{Abishev:2015pqa,Ivashchuk:1999jd}
\begin{align}
    \mathrm{d}{s^2} &= -f(r)\,\d{t^2} + \frac{\d{r^2}}{f(r)} + r^2(H_eH_m)^h\,\d{\Omega_{(2)}^2}\,,\\
    f(r) &= (H_eH_m)^{-h}\left(1 - \frac{2\xi}{r}\right)\,,\\
    e^{-2\lambda\phi} &= \left(\frac{H_e}{H_m}\right)^{2-h}\,,\\
    F_{(2)} &= \frac{q_e}{r^2}H_e^{-2}H_m^{2-2h}\,\d{t}\wedge\d{r} + q_m\sin{\theta}\,\d{\theta}\wedge\d{\varphi}\,,
\end{align}
where the functions $H_\alpha(r)$ ($\alpha=e,m$ and $\overline{\alpha}=m,e$) satisfy
\begin{equation}\label{eq:HaEqn}
    r^2\de{}{r}\left[r^2\left(1 - \frac{2\xi}{r}\right)\frac{H_\alpha'(r)}{H_\alpha(r)}\right] = - h^{-1}q_\alpha^2H_\alpha^{-2}H_{\overline{\alpha}}^{2-2h} \,.
\end{equation}
Imposing the boundary conditions $H_\alpha\to1$ for $r\to\infty$ and $H_\alpha>0$ for $r\to2\xi$, this solution is well-behaved outside the outer horizon, $r>2\xi$. For convenience we have set $\phi_\infty=0$; a constant shift to $\phi$ is compensated for by a rescaling of the electric and magnetic charges.

While closed-form solutions of equation~\eqref{eq:HaEqn} do not exist for general $\lambda$, one can show that series solutions of the form
\begin{equation}
    H_\alpha(r) = 1 + \frac{P_\alpha^{(1)}}{r} + \frac{P_\alpha^{(2)}}{r^2} + \frac{P_\alpha^{(3)}}{r^3} + \cdots
\end{equation}
always converge on $(2\xi,\infty)$ \cite{Ivashchuk:1999jd}. In addition, there is a first-integral of equation~\eqref{eq:HaEqn} which when evaluated at $r\to\infty$ gives \cite{Abishev:2015pqa}
\begin{equation}\label{eq:FirstInt}
    (P_e^{(1)})^2 + (P_m^{(1)})^2 + 2(h-1)P_e^{(1)}P_m^{(1)} + 2\xi(P_e^{(1)} + P_m^{(1)}) - h^{-1}(q_e^2+q_m^2) = 0\,.
\end{equation}
Along with equation~\eqref{eq:HaEqn}, this allows one to solve for the coefficients $P_\alpha^{(k)}$ order-by-order in terms of $P_\alpha^{(1)}$, $\xi$, $h$ and the $q_\alpha$ alone. The entire series is then fixed by prescribing a physical parameter, such as the mass or temperature. In section~\ref{sec:MassEntropyCorrections} we will focus only on those cases where the functions $H_\alpha$ take a particularly manageable form.\footnote{With this parametrization the Einstein-Maxwell ($h=2$) solution is $H_e(r)=\left(1+\frac{P}{r}\right)^\frac{q_e^2}{q_e^2+q_m^2}$ and $H_m(r)=\left(1+\frac{P}{r}\right)^\frac{q_m^2}{q_e^2+q_m^2}$, which takes the usual Reissner-Nordstr\"om form after the change of coordinates $r\to r-P$. Note that only the product $H_eH_m=1+\frac{P}{r}$ appears in the metric and field strength.}

\subsection{Physical Parameters and Extremality}
\label{sec:Extremality}

In writing down the solution to the equations of motion we have introduced the parameters $\xi$, $P_e^{(k)}$ and $P_m^{(k)}$; we would like to interpret these in terms of the physical properties of the black hole. Ultimately the full solution may be determined by specifying only the black hole's charges and temperature.

Metric singularities occur at $r=0$ and $r=2\xi$, which set the locations of the horizons. Extremality thus corresponds to the limit $\xi\to0$, and requiring the regularity of the Euclidean section at the outer horizon gives the black hole temperature as
\begin{equation}\label{eq:EMdTemp}
    T = \frac{f'(2\xi)}{4\pi} = \frac{1}{8\pi\xi}\big[H_e(2\xi)H_m(2\xi)\big]^{-h} \,.
\end{equation}
The areas of the inner and outer horizons are
\begin{equation}\label{eq:EMdArea}
\begin{aligned}
    A_- &= \lim_{r\to0^+}4\pi r^2\big[H_e(r)H_m(r)\big]^h\,,\\
    A_+ &= 16\pi\xi^2\big[H_e(2\xi)H_m(2\xi)\big]^h\,.
\end{aligned}
\end{equation}
The Hawking-Bekenstein entropy, $S=A_+/4G_\text{N}=2\pi A_+$, thus vanishes in the extremal limit only when $A_+\to A_-=0$. Using equations~\eqref{eq:EMdTemp} and \eqref{eq:EMdArea} one finds that the temperature and entropy are related according to
\begin{equation}
    TS = 4\pi\xi\,,
\end{equation}
so that at least one of $T$ and $S$ must vanish at extremality in the classical solution.

In discussing the thermodynamics of these black holes we will need the electric and magnetic potentials which are conjugate to their respective charges. Using equation~\eqref{eq:HaEqn}, the 4-potential is
\begin{equation}
    A_{(1)} = \left[\frac{h}{q_e}r^2\left(1 - \frac{2\xi}{r}\right)\frac{H_e'(r)}{H_e(r)} + A_\text{h}\right]\,\d{t} + q_m(W-\cos{\theta})\,\d{\varphi}\,,
\end{equation}
where the constants $A_\text{h}$ and $W$ may be fixed by a gauge choice. From this we may read off the black hole's (gauge-invariant) electric potential and the analogous magnetic potential relative to infinity,
\begin{equation}
    \Phi \equiv A_t\big|_{2\xi} - A_t\big|_\infty = \frac{hP_e^{(1)}}{q_e} \,, \qquad \Psi = \frac{hP_m^{(1)}}{q_m} \,.
\end{equation}
Although the factors of $q_\alpha$ seem misplaced, recall that the $P_\alpha^{(1)}$ are functions of the $q_\alpha$ as well. By expanding the metric solution at infinity one finds that the gravitational mass is
\begin{equation}
    M = 4\pi\big[2\xi+h(P_e^{(1)}+P_m^{(1)})\big] \,.
\end{equation}
Equation~\eqref{eq:FirstInt} along with the dilatonic charge, defined via $\phi=q_\phi/r+\cdots$, allows one to write the mass as
\begin{equation}
    M^2 = (8\pi\xi)^2 + 2\big(Q_e^2+Q_m^2-Q_\phi^2\big)\,,
\end{equation}
where we have introduced the rescaled charges $Q_\alpha=4\pi q_\alpha$; we will use both $q_\alpha$ and $Q_\alpha$ throughout the remaining sections. Very nicely, the above expression for the mass is independent of the coupling $\lambda$. Note, however, that $Q_\phi=-4\pi h\lambda(P_e^{(1)}-P_m^{(1)})$ is not an independent parameter; in particular, one always has $M^2>0$. In the extremal limit $\xi\to0$ the WGC bound~\eqref{eq:WGCcharges} is saturated, but it remains to be seen whether higher-derivative corrections will decrease the mass and lead to a strict inequality, or increase the mass and lead to such black holes \textit{not} satisfying the bound.

To have a controlled classical solution on which to consider higher-derivative corrections we should require $T<\infty$ and $S>0$ at extremality. Pure electric or magnetic black holes have vanishing area for any $h\neq2$, and in fact have diverging temperature if $h<1$, so that we will have to treat these cases with care. The curiosity of infinite-temperature black holes was pursued in \cite{Holzhey:1991bx}, where it was argued that one should morally think of these solutions as elementary particles and that while the temperature is formally infinite, the rate of thermal radiation emission does go to zero as a result of an infinite mass gap. A qualitative difference between the $h<1$ and $h>1$ regimes for single-charge black holes was also found in \cite{Horowitz:2019eum} when considering connections between the weak gravity conjecture and cosmic censorship.

\section{Euclidean Action and Thermodynamics}
\label{sec:ThermoOutline}

\subsection{Outline of General Procedure}
\label{sec:GeneralProcedure}
Thermodynamic ideas have recently been pursued in calculating corrections to extremal black holes' charge-to-mass or angular momentum-to-mass ratio \cite{Reall:2019sah,Cheung:2019cwi,Cheung:2018cwt}. Here we outline how one may use the Euclidean action to compute thermodynamic quantities, and in particular their higher-derivative corrections, for any finite temperature black hole solution. For a recent thorough discussion, see \cite{Reall:2019sah}. We are ultimately interested in extremal black holes, and so when executing the described procedure we will find it convenient to work with the temperature as an expansion parameter.

Following the usual procedure, temporarily restrict attention to $r<R$ when making a Wick rotation to Euclidean time, $(it)\sim (it)+\beta$, with $\beta=1/T$ the inverse temperature. Eventually the $R\to\infty$ limit will be taken. The action in equation~\eqref{eq:EMdAction} does not lead to a well-posed variational problem; boundary terms are required to ensure that the action is stationary under all variations of the metric and vector potential which vanish on $\partial\mathcal{M}$. The appropriate choice is to include a Gibbons-Hawking-York (GHY) term, which leads to
\begin{equation}
    I_{\text{E},0}^R[g,A,\phi] = -\int_\mathcal{M}\d[4]{x}\,\sqrt{g}\left[\frac{1}{2}R - \frac{1}{2}(\partial\phi)^2 - \frac{1}{4}e^{-2\lambda\phi}\big(F^2\big)\right] - \oint_{\partial\mathcal{M}}\d[3]{\Sigma}\,\sqrt{h}\big(\mathcal{K}-\mathcal{K}_0\big)\,,
\end{equation}
where $\mathcal{K}$ is the trace of the extrinsic curvature on $\partial\mathcal{M}$. Since the GHY term at $r=R$ diverges in the infinite-volume limit we have subtracted off the analogous quantity for flat space, $\mathcal{K}_0$, to regularize the action and ensure that it remains finite in the infinite-volume limit.

The action $I_{\text{E},0}^R$ is invariant under all variations of the metric which vanish on $\partial\mathcal{M}$. In addition, the appropriate boundary conditions for the Maxwell field are to prescribe $A_\|$ on the boundary, i.e.~fix the electric potential at the horizon and the magnetic charge. This may be seen by rewriting the Maxwell term as
\begin{equation}
\begin{aligned}
    -\int_\mathcal{M}\d[4]{x}\,\sqrt{g}\left[-\frac{1}{4}e^{-2\lambda\phi}\big(F^2\big)\right] &= -\frac{1}{2}\int_\mathcal{M}\d[4]{x}\,\sqrt{g}\Big[\nabla_\mu\big(e^{-2\lambda\phi}F^{\mu\nu}\big)A_\nu\big]\\
    &\hspace{80pt} {}+\frac{1}{2}\oint_{\partial\mathcal{M}}\d[3]{\Sigma}\,\sqrt{h}\big(n_\mu e^{-2\lambda\phi}F^{\mu\nu}A_\nu\big) \,.
\end{aligned}
\end{equation}
The surface term is invariant under variations of the 4-potential which vanish at the horizon, which amounts to fixing $A_t$ and $A_\varphi$. Thus when working directly with the Euclidean action it is appropriate to work in the grand canonical ensemble, in which the temperature, electric potential and magnetic charge are the independent quantities \cite{Hawking:1995ap,Gibbons:1976ue}.

After having regularized the boundary term at $r=R$ we may safely take the infinite-volume limit and consider
\begin{equation}
    I_{\text{E},0}[g,A,\phi] \equiv \lim_{R\to\infty}I_{\text{E},0}^R[g,A,\phi]\,.
\end{equation}
The GHY term at infinity contributes to the Euclidean action even in this limit \cite{Gibbons:1976ue}. Evaluating on the EMd solution ($\overline{g}$, $\overline{A}$, $\overline{\phi}$) gives
\begin{equation}
    I_{\text{E},0}[\overline{g},\overline{A},\overline{\phi}] = 4\pi\beta\big(\xi + hP_m^{(1)}\big) = \frac{\beta}{2}\big(M - Q_e\Phi + Q_m\Psi\big)\,.
\end{equation}
The Smarr-like formula, $M=2TS+Q_e\Phi+Q_m\Psi$, implies that we may write
\begin{equation}\label{eq:GibbsDef}
    I_\text{E} = \beta G\equiv \beta(M-TS-Q_e\Phi) \,,
\end{equation}
where $G$ is the free energy. Importantly, this relationship between the free energy and Euclidean action remains true even when higher-derivative corrections are included, as long as $S$ is taken to be the Wald entropy \cite{Wald:1993nt}. Using the first law,
\begin{equation}\label{eq:FirstLaw}
\begin{aligned}
    \d{M} &= T\,\d{S} + \Phi\,\d{Q_e} + \Psi\,\d{Q_m} \,,\\
    \d{G} &= -S\,\d{T} - Q_e\,\d{\Phi} + \Psi\,\d{Q_m} \,,
\end{aligned}
\end{equation}
we may read off
\begin{equation}\label{eq:ThermoDefs}
    S = -\left(\del{G}{T}\right)_{\Phi,Q_m} \,, \qquad Q_e = -\left(\del{G}{\Phi}\right)_{T,Q_m} \,, \qquad \Psi = \left(\del{G}{Q_m}\right)_{T,\Phi} \,.
\end{equation}
These relations allow one to compute thermodynamic quantities in the presence of higher-derivative corrections, to which we turn next.

\subsection{Higher-Derivative Corrections}
\label{sec:HDCorrections}
Effective theories allow one to systematically parametrize the effects of UV physics in terms of a small number of numerical coefficients. The values of these Wilson coefficients are determined by the UV theory, up to field redefinitions.

In string frame we assume an action of the form
\begin{equation}
    I = \int\d[4]{x}\,\sqrt{-g}\,e^{-2\lambda\phi}\big[\mathcal{L}_0(g,A,\partial\phi) + \mathcal{L}_\text{h.d.}(g,A,\partial\phi)\big] \,,
\end{equation}
this structure being motivated by the low-energy string effective action at leading order in $g_s$. Returning to Einstein frame, the most general collection of parity-preserving, four-derivative terms for 4D EMd theory may be written as
\begin{align}\label{eq:hdAction}
    \alpha_iI_i &\equiv \int\d[4]{x}\,\sqrt{-g}\left[\frac{\alpha_1}{4}e^{-6\lambda\phi}\big(F^2\big)^2 + \frac{\alpha_2}{4}e^{-6\lambda\phi}\big(F\widetilde{F}\big)^2 + \frac{\alpha_3}{2}e^{-4\lambda\phi}\big(FFW\big) + \frac{\alpha_4}{2}e^{-2\lambda\phi}(R_\text{GB}) \right. \notag\\
    &\qquad\qquad \left. {} + \frac{\alpha_5}{4}e^{-2\lambda\phi}(\partial\phi)^4 + \frac{\alpha_6}{4}e^{-4\lambda\phi}(\partial\phi)^2\big(F^2\big) + \frac{\alpha_7}{4}e^{-4\lambda\phi}(\partial\phi\partial\phi FF) \right] \,,
\end{align}
where we have used the compact notation
\begin{equation}
\begin{aligned}
    (FFW) &= F_{\mu\nu}F_{\rho\sigma}W^{\mu\nu\rho\sigma} \,, & (R_\text{GB}) &= R_{\mu\nu\rho\sigma}R^{\mu\nu\rho\sigma} - 4R_{\mu\nu}R^{\mu\nu} + R^2 \,,\\
    (\partial\phi)^4 &= (\partial_\mu\phi\partial^\mu\phi)^2 \,, & (\partial\phi\partial\phi FF) &= \partial_\mu\phi\partial_\nu\phi F^\mu{}_\rho F^{\nu\rho} \,,
\end{aligned}
\end{equation}
and where $W$ is the Weyl tensor. We have chosen to parametrize these in terms of $R_\text{GB}$, rather than $W^2$, so that we may more easily compare with previous Einstein-Maxwell results in the $\lambda\to0$ limit where the Gauss-Bonnet term becomes topological. Note that the first duality of equation~\eqref{eq:Dualities} is maintained, while the electromagnetic duality is broken. We will treat electric and magnetic solutions separately in section~\ref{sec:MassEntropyCorrections}.

Upon adding these higher-derivative terms the equations of motion which need to be solved are altered, leading to perturbed solutions for $g$, $A$ and $\phi$. We favor, however, a \textit{thermodynamic} approach which has recently been used in \cite{Reall:2019sah} and \cite{Cheung:2019cwi}. Equations~\eqref{eq:GibbsDef} and \eqref{eq:ThermoDefs} are used along with the following fact:
\begin{equation}\label{eq:IEequiv}
    I_\text{E}[g,A,\phi] = I_\text{E}[\overline{g},\overline{A},\overline{\phi}] + \mathcal{O}(\alpha_i^2) \,.
\end{equation}
We emphasize that $\overline{g}$, $\overline{A}$ and $\overline{\phi}$ are the EMd solutions without higher derivative corrections, whereas
$I_\text{E}$ is the full Euclidean action including the higher-derivative terms. For a proof of the above in Einstein gravity we refer the reader to Ref.~\cite{Reall:2019sah}.

\begin{figure}[t]
    \centering
    \begin{tikzpicture}
        \node[block, text width=8em] at (0,1.5) (IE) {Euclidean Action};
        \node[block, text width=9em] at (0,0) (GCE) {\textbf{Grand Canonical}\\ Fixed: $T,\Phi,Q_m$};
        \node[block, text width=9em] at (5,0) (CE) {\textbf{Canonical}\\ Fixed: $T,Q_e,Q_m$};
        \node[block, text width=9em] at (10,0) (MCE) {\textbf{Microcanonical}\\ Fixed: $M,Q_e,Q_m$};
        \node[block, text width=7em] at (5,-1.5) (WGC) {WGC: $z_\text{ext}>1$};
        \node[block, text width=9em] at (10,-1.5) (dS>0) {WGC: $\Delta S\big|_{z=1}>0$};
        \path[line] (IE) -- (GCE);
        \path[line] (GCE) -- (CE);
        \path[line] (CE) -- (MCE);
        \path[line] (CE) -- (WGC);
        \path[line] (MCE) -- (dS>0);
    \end{tikzpicture}
    \caption{Relationship between different ensembles and the weak gravity conjecture.}
    \label{fig:ensembleFlowChart}
\end{figure}
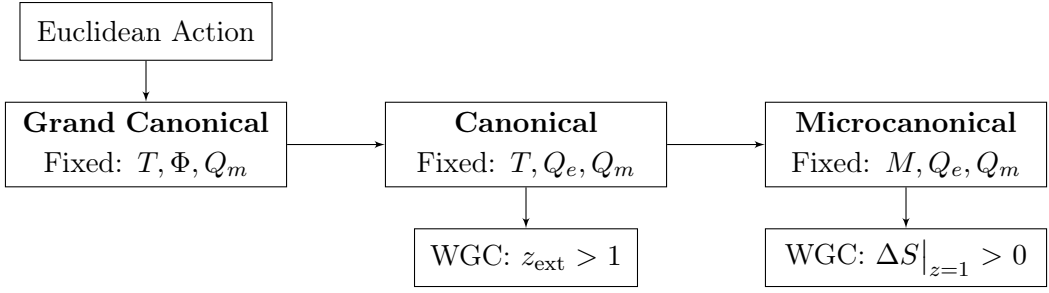

Given equations~\eqref{eq:ThermoDefs} and \eqref{eq:IEequiv}, corrections in the grand canonical ensemble are
\begin{align}
    \left(\del{G}{\alpha_i}\right)_{T,\Phi,Q_m} &= TI_{\text{E},i} \,,\\
    \left(\del{S}{\alpha_i}\right)_{T,\Phi,Q_m} &= -\left(\del{(TI_{\text{E},i})}{T}\right)_{\Phi,Q_m,\alpha_j} \,,\\
    \left(\del{Q_e}{\alpha_i}\right)_{T,\Phi,Q_m} &= -T\left(\del{I_{\text{E},i}}{\Phi}\right)_{T,Q_m,\alpha_j} \,,\\
    \left(\del{\Psi}{\alpha_i}\right)_{T,\Phi,Q_m} &= T\left(\del{I_{\text{E},i}}{Q_m}\right)_{T,\Phi,\alpha_j} \,,\\
    \left(\del{M}{\alpha_i}\right)_{T,\Phi,Q_m} &= TI_{\text{E},i} - T\left(\del{(TI_{\text{E},i})}{T}\right)_{\Phi,Q_m,\alpha_j} -T\Phi\left(\del{I_{\text{E},i}}{\Phi}\right)_{T,Q_m,\alpha_j}\,,
\end{align}
where $I_{E,i}$ denote the Euclidean versions of the terms in the four-derivative action, equation~\eqref{eq:hdAction}. One may then transition to different ensembles, as outlined in figure~\ref{fig:ensembleFlowChart}, by inverting to leading order in $\alpha_i$. The weak gravity conjecture is directly a statement about the charge-to-mass ratio in the canonical ensemble, where at fixed temperature ($T=0$) we expect $Q/M\sim z_\text{ext}>1$.

One also expects that at fixed mass and charge, the entropy of $z=1$ black holes increases in the presence of higher derivative corrections. Using standard thermodynamic manipulations, the corrections to the entropy in the microcanonical ensemble are\footnote{The first equality is found by using the definition of the free energy, equation~\eqref{eq:FirstLaw} and the triple product rule along with \[\left(\del{S}{\alpha_i}\right)_{T,\Phi,Q_m} = \left(\del{S}{M}\right)_{Q_e,Q_m,\alpha_i}\left(\del{M}{\alpha_i}\right)_{T,\Phi,Q_m} + \left(\del{S}{Q_e}\right)_{M,Q_m,\alpha_i}\left(\del{Q_e}{\alpha_i}\right)_{T,\Phi,Q_m} + \left(\del{S}{\alpha_i}\right)_{M,Q_e,Q_m} \,. \]}
\begin{equation}\label{eq:EntropyAction}
    \left(\del{S}{\alpha_i}\right)_{M,Q_e,Q_m} = -\left(\del{(\beta G)}{\alpha_i}\right)_{T,\Phi,Q_m} = -I_{\text{E},i}(T,\Phi,Q_m)\,,
\end{equation}
where $T$ and $\Phi$ are implicitly functions of $M$, $Q_e$ and $Q_m$. We find that these corrections are never $\mathcal{O}(\alpha_i)$ for extremal black holes, in agreement with the Supplemental Material of~\cite{Hamada:2018dde}.

\section{Mass and Entropy Corrections}
\label{sec:MassEntropyCorrections}
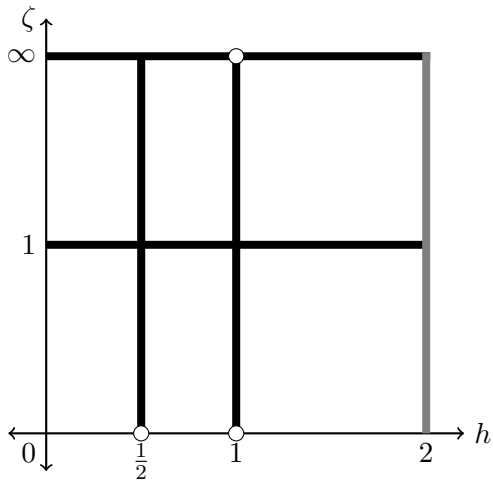
\begin{figure}[t]
    \centering
    \newcommand{\s}{5}
    \newcommand{\lw}{3}
    \begin{tikzpicture}
        \draw[<->,thick] (-0.1*\s,0) -- (1.1*\s,0);
        \draw[<->,thick] (0,-0.1*\s) -- (0,1.1*\s);
        \draw[line width=\lw] (0,\s) -- (\s+0.0175*\lw,\s);
        \draw[line width=\lw] (0.5*\s,0) -- (0.5*\s,\s);
        \draw[line width=\lw] (0.25*\s,0) -- (0.25*\s,\s);
        \draw[line width=\lw] (0,0.5*\s) -- (\s,0.5*\s);
        \draw[line width=\lw,gray] (\s,0) -- (\s,\s+0.0175*\lw);
        \draw[fill=white] (0.5*\s,0) circle (0.1);
        \draw[fill=white] (0.5*\s,\s) circle (0.1);
        \draw[fill=white] (0.25*\s,0) circle (0.1);
        \node[right] at (1.1*\s,0) {$h$};
        \node[left] at (0,1.1*\s) {$\zeta$};
        \node[below] at (\s,0) {$2$};
        \node[below] at (0.5*\s,0) {$1$};
        \node[below] at (0.25*\s,0) {$\frac{1}{2}$};
        \node[left] at (0,\s) {$\infty$};
        \node[left] at (0,0.5*\s) {$1$};
        \node[below left] at (0,0) {$0$};
    \end{tikzpicture}
    \caption{Charge/dilaton coupling parameter space for black holes with a massless dilaton. The black lines denote those regions covered by sections~\ref{sec:Magnetic}, \ref{sec:EqualCharge}, \ref{sec:h=1} and \ref{sec:h=1/2}. The Einstein-Maxwell case is shown in gray.}
    \label{fig:paramSpace}
\end{figure}

We turn now to the execution of the procedure outlined in section~\ref{sec:ThermoOutline}. There are several special choices of the exponential coupling constant and/or charges for which equation~\eqref{eq:HaEqn} is exactly solvable. We will use five such closed-form solutions of equation~\eqref{eq:HaEqn} which allow us to calculate corrections analytically; (i) pure magnetic, (ii) pure electric, (iii) equal-charge, (iv) dyonic with string theory coupling ($h=1$), and (v) dyonic with 5D KK reduction coupling ($h=1/2$). Corrections to the mass for the case of the magnetically-charged Garfinkle-Horowitz-Strominger black hole ($h=1$) have been computed in \cite{Natsuume:1994hd}. See figure~\ref{fig:paramSpace} for the complete parameter space.

In presenting the following consistency conditions, it is worth noting that $\alpha_3$ and $\alpha_4$ are small when gravitational effects are negligible, and that unitarity of the $S$-matrix requires that $\alpha_1$, $\alpha_2$, $\alpha_5$ and $\alpha_7$ are all positive. We will discuss these statements further in section~\ref{sec:WGC}.

It is convenient to introduce the following notation when discussing corrections in the canonical ensemble:
\begin{equation}
    X_0 \equiv X\big|_{\alpha_j=0} \,, \qquad \overline{\delta}_iX \equiv \left.\frac{1}{X}\left(\del{X}{\alpha_i}\right)_{T,Q_e,Q_m}\right|_{\alpha_j=0} \,, \qquad \zeta\equiv\frac{Q_m}{Q_e} \,.
\end{equation}
In this section we present only the thermodynamics of the EMd solutions and those corrections which are relevant for the weak gravity conjecture.\footnote{Complete corrections are available upon request.} For clarity we work only with positive charges; negative charges may be accounted for with the addition of appropriate absolute values.

\subsection{Magnetic}
\label{sec:Magnetic}
The simplest case for our purpose is the magnetically charged black hole, as here $\Phi=Q_e=0$ and what we call the grand canonical and canonical ensembles are one and the same. With $Q_e=0$, equation~\eqref{eq:HaEqn} has solutions
\begin{equation}
    H_e(r) = 1\,, \qquad H_m(r) = 1 + \frac{P_m}{r} \,,
\end{equation}
where $hP_m(P_m+2\xi)=q_m^2$. This classical solution has area, temperature and horizon dilaton at extremality given by
\begin{equation}
    A_- = 0\,, \qquad T\to\begin{cases}
    0 & h>1\\
    \infty & h<1
    \end{cases} \,,\qquad e^{-2\lambda\phi} \sim \left(\frac{r}{q_m}\right)^{2-h} \to 0\,.
\end{equation}
Despite having vanishing classical area, the derivative expansion is well-behaved due to the suppression from higher-and-higher powers of $e^{-2\lambda\phi}$. For example, terms of the form $e^{-(2+4k)\lambda\phi}(F^2)^{k+1}$ have an expansion parameter
\begin{equation}
    e^{-4\lambda\phi}(F^2) \sim \frac{1}{q_m^2}-\frac{2r}{q_m^3} + \cdots
\end{equation}
at extremality near the horizon. As long as the magnetic charge is large enough, the derivative expansion is under control.

While the interpretation of infinite temperature in the classical solution is suspect, we note that the regions $h>1$ and $h<1$ are treated identically in the following thermodynamic approach. Indeed, defining $\tau^{h-1}\equiv Q_mT/\sqrt{h}$, the EMd solution has
\begin{align}
    G_0 &= \sqrt{h}Q_m\left[1 - \frac{h-1}{2h}\tau - \frac{1}{8}\tau^2 + \cdots\right] \,,\\
    S_0 &= \frac{Q_m^2\tau^{2-h}}{2h}\left[1 + \frac{h}{2(h-1)}\tau + \cdots\right] \,,\\
    \Psi_0 &= \sqrt{h}\left[1 - \frac{1}{2}\tau - \frac{h+1}{8(h-1)}\tau^2 + \cdots\right] \,,\\
    M_0 &= \sqrt{h}Q_m\left[1 + \frac{2-h}{2h}\tau + \frac{3-h}{8(h-1)}\tau^2 + \cdots\right] \,.
\end{align}
The extremal limit is $\tau\to0$ for all $h\neq1$. For $h=1$ the series expansions break down and extremality becomes $T\to Q_m^{-1}$. In the canonical ensemble, we find
\begin{subequations}
\begin{align}
    \overline{\delta}_1M &= - \frac{h^2}{5q_m^2}\left[1 + \frac{(2-h)(4h+1)}{2h(h-1)}\tau + \cdots\right] \,,\\
    \overline{\delta}_3M &= -\frac{h^2}{30q_m^2}\bigg[(11-4h)\\
    &\qquad\qquad \left.+ \frac{(2-h)[873-875h+202h^2-60(2-h)(2h-3)\log{\tau}]}{6h(h-1)}\tau + \cdots\right] \,, \notag\\
    \overline{\delta}_4M &= -\frac{h^2(2-h)}{10q_m^2}\bigg[(5-h)\\
    &\left.- \frac{720-3500h+4432h^2-1943h^3+286h^4-60h(2-h)(8-9h+2h^2)\log{\tau}}{12h^2(h-1)}\tau + \cdots\right] \,, \notag\\
    \overline{\delta}_5M &= -\frac{h^2(2-h)^2}{80q_m^2}\left[1 + \frac{6-349h+143h^2-60h(2-h)\log{\tau}}{6h(h-1)}\tau + \cdots\right] \,, \\
    \overline{\delta}_6M &= -\frac{h^2(2-h)}{20q_m^2}\left[1 + \frac{12-268h+101h^2-60h(2-h)\log{\tau}}{12h(h-1)}\tau + \cdots\right] \,,
\end{align}
\end{subequations}
and $\overline{\delta}_2M=\overline{\delta}_7M=0$. In the extremal limit the charge-to-mass ratio, $z=\sqrt{h}Q_m/M$, is
\begin{equation}
    z_\text{ext} = 1 + \frac{h^2}{10q_m^2}\left[2\alpha_1 + \frac{1}{3}(11-4h)\alpha_3 + (2-h)(5-h)\alpha_4 + \frac{1}{8}(2-h)^2\alpha_5+\frac{1}{2}(2-h)\alpha_6\right] \,.
\end{equation}
Note that while smooth at $h=1$, this result does not directly apply for $h=1$ since there our expansion in $\tau$ is ill-defined. For $h\neq1$, the weak gravity conjecture is then
\begin{equation}\label{eq:Cmag}
    \mathcal{C}_\text{mag}^{h\neq1}(\alpha_i) \equiv 2\alpha_1 + \frac{1}{3}(11-4h)\alpha_3+(2-h)(5-h)\alpha_4+\frac{1}{8}(2-h)^2\alpha_5+\frac{1}{2}(2-h)\alpha_6 > 0 \,.
\end{equation}
Note that as functions of $h$ the coefficients of each $\alpha_i$ are of definite sign, with the $\alpha_1$ and $\alpha_5$ contributions to $\mathcal{C}_\text{mag}^{h\neq1}$ always being positive. As a check, this reduces to the Einstein-Maxwell condition for magnetically charged black holes for $\lambda\to0$:
\begin{equation}
    \mathcal{C}_\text{mag}^{h\neq1}(\alpha_i)\big|_{h=2} = 2\alpha_1 + \alpha_3 > 0 \,.
\end{equation}
To obtain the entropy in the microcanonical ensemble, we must invert
\begin{align}
    M &= \sqrt{h}Q_m\left(1 - \frac{h^2}{10q_m^2}\mathcal{C}_\text{mag}^{h\neq1}(\alpha_i) + \frac{2-h}{2h}\tau + \cdots\right)\,.
\end{align}
in favor of $\tau$. When $z=\sqrt{h}Q_m/M=1$, we find
\begin{equation}
    \tau \approx \frac{h^3\,\mathcal{C}_\text{mag}^{h\neq1}(\alpha_i)}{5(2-h)q_m^2} \,,
\end{equation}
which leads to\footnote{If $h=2$, then one goes to the next order in $\tau$ and finds $\tau\sim\sqrt{a_i}$. There would then be an $\mathcal{O}(\sqrt{\alpha_1})$ correction to the entropy, as in \cite{Hamada:2018dde}.}
\begin{equation}
    \Delta S\big|_{z=1} = \frac{Q_m^2}{2h}\left(\frac{h^3\,\mathcal{C}_\text{mag}^{h\neq1}(\alpha_i)}{5(2-h)q_m^2}\right)^{2-h}\big[ 1 + \mathcal{O}(\alpha_i) \big] \,.
\end{equation}
This leading shift to the entropy comes entirely from evaluating $S_0$ at nonzero temperature.

\subsection{Electric}
\label{sec:Electric}
For a massless dilaton the classical solution has
\begin{equation}
    H_e(r) = 1 + \frac{P_e}{r} \,,\qquad H_m(r) = 1 \,,
\end{equation}
where $hP_e(P_e+2\xi)=q_e^2$. The area, temperature and horizon dilaton at extremality are
\begin{equation}
    A_- = 0 \,, \qquad T\to\begin{cases}
    0 & h>1\\
    \infty & h<1
    \end{cases}\,,\qquad e^{-2\lambda\phi} \sim\left(\frac{q_e}{r}\right)^{2-h} \to\infty\,.
\end{equation}
In contrast with the magnetic case, the diverging dilaton spoils the derivative expansion, since higher-derivative operators are enhanced near the outer horizon, e.g.
\begin{equation}
    e^{-10\lambda\phi}\big(F^2\big)^3 \gg e^{-6\lambda\phi}\big(F^2\big)^2 \,.
\end{equation}
This divergence may be avoided by stabilizing the dilaton with a mass $m_\phi\gtrsim\frac{|\lambda|}{M}$ (for masses below this the solution approaches that of a massless dilaton near the outer horizon and the divergences survive). The classical solution now takes the form~\cite{Gregory:1992kr,Horne:1992bi}
\begin{align}
    \d{s^2} &= -f(r)\,\d{t^2} + \frac{\d{r^2}}{f(r)} + R(r)^2\,\d{\Omega^2} \,,\\
    f(r) &= \left(1 - \frac{M}{4\pi r} + \frac{q_e^2}{2r^2}\right) - \frac{\lambda^2q_e^4}{10m_\phi^2r^6} + \cdots \,,\\
    R(r) &= r\left(1 - \frac{\lambda^2q_e^4}{7m_\phi^4r^8}+\cdots\right) \,,\\
    \phi &= -\frac{\lambda^2q_e^2}{m_\phi^2r^4} + \cdots \,,\\
    F_{tr} &= \frac{q_e}{r^2}\left(1 - \frac{2\lambda^2q_e^2}{m_\phi^2r^4} + \cdots\right) \,.
\end{align}
We have checked that running the thermodynamic procedure reproduces the same leading-order corrections as integrating out $\phi$ at tree-level. The Wilson coefficients for the resulting Einstein-Maxwell theory are
\begin{equation}
    \alpha_1' = \alpha_1 + \frac{\lambda^2}{2m_\phi^2} + \mathcal{O}\Big(\frac{1}{m_\phi^4}\Big) \,,\qquad \alpha_{2,3,4}' = \alpha_{2,3,4} + \mathcal{O}\Big(\frac{1}{m_\phi^4}\Big) \,,\qquad \alpha_{5,6,7}' = 0 \,.
\end{equation}
We may thus immediately write down the charge-to-mass ratio, $z=\sqrt{2}Q_e/M$, at extremality,
\begin{equation}
    z_\text{ext} = 1 + \frac{2}{5q_e^2}\big(2\alpha_1'-\alpha_3'\big) = 1 + \frac{2}{5q_e^2}\big(2\alpha_1-\alpha_3\big) + \frac{2\lambda^2}{5q_e^2m_\phi^2} + \cdots \,.
\end{equation}
The weak gravity conjecture is then
\begin{equation}
    \mathcal{C}_\text{el}(\alpha_i;m_\phi) \equiv 2\alpha_1'-\alpha_3' = 2\alpha_1-\alpha_3+\frac{\lambda^2}{m_\phi^2} > 0 \,,
\end{equation}
and the entropy is corrected as
\begin{equation}
    \Delta S\big|_{z=1} = \frac{4\pi Q_e}{\sqrt{5}}\sqrt{\mathcal{C}_\text{el}(\alpha_i;m_\phi)} + \mathcal{O}(\alpha_i,m_\phi^{-1}) \,.
\end{equation}

\subsection{\texorpdfstring{Dyonic, $Q_e=Q_m$}{Qe=Qm}}
\label{sec:EqualCharge}
For black holes of equal electric and magnetic charges the equations of motion have the following solution:
\begin{equation}
    H_e(r) = H_m(r) = \left(1 + \frac{P}{r}\right)^{1/h} = 1 + \frac{P}{hr} + \frac{(1-h)P^2}{2h^2r^2} + \cdots
\end{equation}
where $P(P+2\xi)=q_e^2=q_m^2\equiv q^2$. In fact, since the dilaton profile is trivial, this is a solution of Einstein-Maxwell theory for which $g_{\mu\nu}$ and $A_\mu$ are both independent of $\lambda$. The usual dyonic, Reissner-Nordstr\"om solution is found after the change of coordinates $r\to r-P$. One is then faced with computing corrections due to only $(F\widetilde{F})^2$ and $R_\text{GB}$ in Einstein-Maxwell theory. The charge-to-mass ratio, $z=2Q/M$, at extremality is simply
\begin{equation}
    z_\text{ext} = 1 + \frac{2\alpha_2}{5q^2} \,,
\end{equation}
so that the weak gravity conjecture is
\begin{equation}\label{Cequal}
    \mathcal{C}_{Q_e=Q_m}(\alpha_i) \equiv \alpha_2 > 0 \,.
\end{equation}
The entropy of an extremal black hole is corrected as
\begin{equation}
    \Delta S\big|_{z=1} = \frac{8\pi Q}{\sqrt{5}}\sqrt{\alpha_2} + \mathcal{O}(\alpha_2,\alpha_4) \,.
\end{equation}
We will use these particularly simple expressions as a cross-check on the remaining two cases.

\subsection{\texorpdfstring{Dyonic, $h=1$}{h=1}}
\label{sec:h=1}
\begin{figure}[t]
    \centering
    \includegraphics[width=0.85\textwidth]{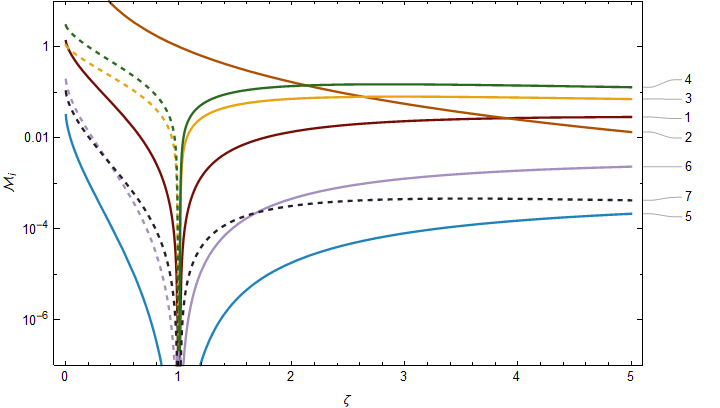}
    \caption{The functions $\mathcal{M}_i(\zeta)$, with solid and dashed lines indicating positive and negative values respectively. Only $\mathcal{M}_2$ is nonzero for $\zeta\to1$.}
    \label{fig:MiZeta}
\end{figure}

For $\lambda^2=1/2$ ($h=1$) the exponential coupling corresponds to that found in the low-energy effective action of string theory. Here we consider dyonic solutions, for which the extremal limit is $T\to0$. Equation~\eqref{eq:HaEqn} has for solutions
\begin{equation}
    H_e(r) = 1 + \frac{P_e}{r} \,, \qquad H_m(r) = 1 + \frac{P_m}{r} \,,
\end{equation}
where $P_\alpha(P_\alpha+2\xi)=q_\alpha^2$. The classical area, temperature and dilaton at the horizon are all well-behaved for extremal solutions:
\begin{equation}
    A_- = 4\pi q_eq_m \,, \qquad T\to 0 \,,\qquad e^{-2\lambda\phi} \to \frac{q_e}{q_m}\,.
\end{equation}
In particular, the derivative expansion is intact as long as both charges are nonzero. The EMd solution has
\begin{align}
    G_0 &= \frac{1-\Phi^2}{2T} + \frac{Q_m^2T}{2(1-\Phi^2)} \,,\\
    S_0 &= \frac{1-\Phi^2}{2T^2} - \frac{Q_m^2}{2(1-\Phi^2)} \,,\\
    Q_{e,0} &= \frac{\Phi}{T} - \frac{Q_m^2\Phi T}{(1-\Phi^2)^2} \,,\\
    \Psi_0 &= \frac{Q_mT}{1-\Phi^2} \,,\\
    M_0 &= \frac{1}{T} - \frac{Q_m^2\Phi^2T}{(1-\Phi^2)^2} \,.
\end{align}
Inverting $Q_{e,0}$ in favor of $\Phi_0$ gives
\begin{align}
    \Phi_0 &= 1 - \frac{1}{2}Q_mT -\frac{1}{8}Q_m(2Q_e+Q_m)T^2 + \cdots \,.
\end{align}
In the canonical ensemble, the above leads to
\begin{align}
    G_0 &= Q_m\left[1 + \frac{1}{8}Q_e^2T^2 + \frac{1}{8}Q_e^2(Q_e+Q_m)T^3 + \cdots\right] \,,\\
    S_0 &= \frac{Q_eQ_m}{2}\left[1 + \frac{1}{2}(Q_e+Q_m)T + \frac{3}{8}(Q_e+Q_m)^2T^2 + \cdots\right] \,,\\
    \Psi_0 &= 1 - \frac{1}{2}Q_eT - \frac{1}{8}Q_e(Q_e+2Q_m)T^2 + \cdots \,,\\
    M_0 &= (Q_e+Q_m)\left[1 + \frac{1}{8}Q_eQ_mT^2 + \frac{1}{8}Q_eQ_m(Q_e+Q_m)T^3 + \cdots\right] \,.
\end{align}
The electromagnetic duality of the classical solution is evident. Corrections to the mass in the canonical ensemble take the form
\begin{equation}
    \overline{\delta}_iM = -\frac{2}{5q_eq_m}\mathcal{M}_i(\zeta) + \mathcal{O}(T^2) \,,
\end{equation}
where
\begin{subequations}\label{eq:ScriptMs}
\begin{align}
    \mathcal{M}_1(\zeta) &= \frac{(1-\zeta)(8+103\zeta-137\zeta^2-37\zeta^3+3\zeta^4)+60\zeta(1-2\zeta^2)\log{\zeta}}{6(1+\zeta)(1-\zeta)^5} \,,\\
    \mathcal{M}_2(\zeta) &= \frac{2}{\zeta^2(1+\zeta)} \,,\\
    \mathcal{M}_3(\zeta) &= -\frac{(1-\zeta)(39-146\zeta-86\zeta^2+334\zeta^3-21\zeta^4)+60\zeta(1-6\zeta+6\zeta^2+\zeta^3)\log{\zeta}}{36(1+\zeta)(1-\zeta)^5} \,,\\
    \mathcal{M}_4(\zeta) &= -\frac{(1-\zeta)(71+111\zeta-309\zeta^2+331\zeta^3-24\zeta^4)+60\zeta(4-6\zeta+4\zeta^2+\zeta^3)\log{\zeta}}{24(1+\zeta)(1-\zeta)^5} \,,\\
    \mathcal{M}_5(\zeta) &= \frac{(1-\zeta)(3+178\zeta+478\zeta^2+178\zeta^3+3\zeta^4)+60\zeta(1+\zeta)(1+5\zeta+\zeta^2)\log{\zeta}}{96(1+\zeta)(1-\zeta)^5} \,,\\
    \mathcal{M}_6(\zeta) &= \frac{(1-\zeta)(9+299\zeta+239\zeta^2-121\zeta^3-6\zeta^4)+60\zeta(2+6\zeta-\zeta^3)\log{\zeta}}{48(1+\zeta)(1-\zeta)^5} \,,\\
    \mathcal{M}_7(\zeta) &= -\frac{5(1-\zeta^2)(1+28\zeta+\zeta^2)+60\zeta(1+3\zeta+\zeta^2)\log{\zeta}}{48(1+\zeta)(1-\zeta)^5} \,.
\end{align}
\end{subequations}
These functions are in fact all finite for $\zeta\to1$, as seen in figure~\ref{fig:MiZeta}.
The charge-to-mass ratio, $z=(Q_e+Q_m)/M$, at extremality is thus
\begin{equation}
    z_\text{ext} = 1 + \frac{2}{5q_eq_m}\alpha_i\mathcal{M}_i(\zeta)\,.
\end{equation}
The weak gravity conjecture for general $\zeta$ is then
\begin{equation}\label{eq:Ch=1}
    \mathcal{C}_\text{dyon}^{h=1}(\alpha_i;\zeta) \equiv \alpha_i\mathcal{M}_i(\zeta) > 0 \,.
\end{equation}
We note that the $\alpha_1$, $\alpha_2$ and $\alpha_5$ contributions to $\mathcal{C}_\text{dyon}^{h=1}$ are always positive, while the $\alpha_7$ contribution is always negative. The equal-charge and magnetic limits agree with those found above (the factor of $4\zeta$ is due to our definition of $\mathcal{M}_i$):
\begin{equation}
    \lim_{\zeta\to1}\mathcal{C}_\text{dyon}^{h=1}(\alpha_i;\zeta) = \mathcal{C}_{Q_e=Q_m}(\alpha_i) \,,\quad \lim_{\zeta\to\infty} (4\zeta)\,\mathcal{C}_\text{dyon}^{h=1}(\alpha_i;\zeta) = \mathcal{C}_\text{mag}^{h\neq1}(\alpha_i)\big|_{h=1} \,.
\end{equation}
The magnetic limit here should not be taken too seriously, since the extremal limit is not captured by the expansions of section~\ref{sec:Magnetic} and there is no reason to expect that the extremal and $h\to1$ limits commute.

For the entropy corrections, inverting
\begin{align}\label{eq:h=1masscorrections}
    M &= (Q_e+Q_m)\left(1 - \frac{2}{5q_eq_m}\mathcal{C}_\text{dyon}^{h=1}(\alpha_i;\zeta) + \frac{1}{8}Q_eQ_m T^2 + \cdots \right)
\end{align}
for $T$ when $z=(Q_e+Q_m)/M=1$ gives
\begin{equation}\label{eq:h=1temp}
    T \approx \frac{16\pi\sqrt{\mathcal{C}_\text{dyon}^{h=1}(\alpha_i)}}{\sqrt{5}\,Q_eQ_m} \,.
\end{equation}
The entropy correction is then
\begin{equation}\label{eq:h=1entropy}
    \Delta S\big|_{z=1} = \frac{4\pi}{\sqrt{5}}(Q_e+Q_m)\sqrt{\mathcal{C}_\text{dyon}^{h=1}(\alpha_i)} + \mathcal{O}(\alpha_i) \,.
\end{equation}
As before, the leading contributions comes only from $\Delta S_0$.

\subsection{\texorpdfstring{Dyonic, $h=1/2$}{h=1/2}}
\label{sec:h=1/2}
With $\lambda^2=3/2$ ($h=1/2$) the exponential coupling corresponds to the KK reduction of Einstein gravity on $\mathcal{M}_4\times S^1$, with the radion playing the role of the dilaton. In this case equation~\eqref{eq:HaEqn} has for solutions
\begin{equation}
    H_e(r) = 1 + \frac{P_e}{r} + \frac{P_e^{(2)}}{r^2} \,, \qquad H_m(r) = 1 + \frac{P_m}{r} + \frac{P_m^{(2)}}{r^2} \,,
\end{equation}
where the coefficients are the positive solutions of
\begin{align}
    2q_\alpha^2 &= \frac{P_\alpha(P_\alpha+2\xi)(P_\alpha+4\xi)}{P_e+P_m+4\xi} \,,\\
    P_\alpha^{(2)} &= \frac{P_eP_m(P_\alpha+2\xi)}{2(P_e+P_m+4\xi)} \,.
\end{align}
The classical area, temperature and horizon value of the dilaton at extremality are
\begin{equation}
    A_- = 4\pi q_eq_m \,, \qquad T\to0 \,,\qquad e^{-2\lambda\phi} \to \frac{q_e}{q_m}\,,
\end{equation}
so that the derivative expansion is well-behaved for non-vanishing charges. Defining $\mathcal{T}\equiv\frac{Q_mT}{\Phi(2\Phi^2-1)}$, the EMd solution has
\begin{align}
    G_0 &= \frac{Q_m\Phi}{\sqrt{2\Phi^2-1}}\left[1 - \frac{1}{2}\mathcal{T} + \frac{\Phi^2}{2}\mathcal{T}^2 + \cdots\right] \,,\\
    S_0 &= \frac{Q_m^2}{2(2\Phi^2-1)^{3/2}}\left[1 - 2\Phi^2\mathcal{T} + \cdots\right] \,,\\
    Q_{e,0} &= \frac{Q_m}{(2\Phi^2-1)^{3/2}}\left[1 - 3\Phi^2\mathcal{T} + \frac{\Phi^2}{2}(1+8\Phi^2)\mathcal{T}^2 + \cdots\right] \,,\\
    \Psi_0 &= \frac{\Phi}{\sqrt{2\Phi^2-1}}\left[ 1 - \mathcal{T} + \frac{3\Phi^2}{2}\mathcal{T}^2 + \cdots\right] \,,\\
    M_0 &= \frac{2Q_m\Phi^3}{(2\Phi^2-1)^{3/2}}\left[1 - \frac{3}{2}\mathcal{T} + \frac{1}{2}(1+3\Phi^2)\mathcal{T}^2 + \cdots\right] \,.
\end{align}
To work in the canonical ensemble it is convenient to introduce
\begin{equation}
    \widetilde{\mathcal{T}} \equiv \frac{(Q_eQ_m)^{2/3}T}{\sqrt{2\big(Q_e^{2/3}+Q_m^{2/3}\big)}}\,.
\end{equation}
With this the EMd solution in the canonical ensemble has
\begin{align}
    G_0 &= \frac{Q_m}{\sqrt{2}}\sqrt{1+\zeta^{-2/3}}\left[1 + \frac{1}{6}\big(1 + 2\zeta^{-2/3}\big)\widetilde{\mathcal{T}}^2 + \cdots\right] \,,\\
    S_0 &= \frac{Q_eQ_m}{2}\left[ 1 + \big( \zeta^{1/3}+\zeta^{-1/3} \big)\widetilde{\mathcal{T}} + \cdots\right] \,,\\
    \Phi_0 &= \frac{1}{\sqrt{2}}\sqrt{1+\zeta^{2/3}}\left[1 - \zeta^{1/3}\widetilde{\mathcal{T}} - \frac{1}{6}\big(5+4\zeta^{2/3}\big)\widetilde{\mathcal{T}}^2 + \cdots\right] \,,\\
    \Psi_0 &= \frac{1}{\sqrt{2}}\sqrt{1+\zeta^{-2/3}}\left[1 - \zeta^{-1/3}\widetilde{\mathcal{T}} - \frac{1}{6}\big(5+4\zeta^{-2/3}\big)\widetilde{\mathcal{T}}^2 + \cdots\right] \,,\\
    M_0 &= \frac{1}{\sqrt{2}}(Q_e^{2/3}+Q_m^{2/3})^{3/2}\left[1 + \frac{1}{2}\widetilde{\mathcal{T}}^2 + \cdots\right] \,.
\end{align}
Again, electromagnetic duality is manifest in the classical solution.

\begin{figure}[t]
    \centering
    \includegraphics[width=0.85\textwidth]{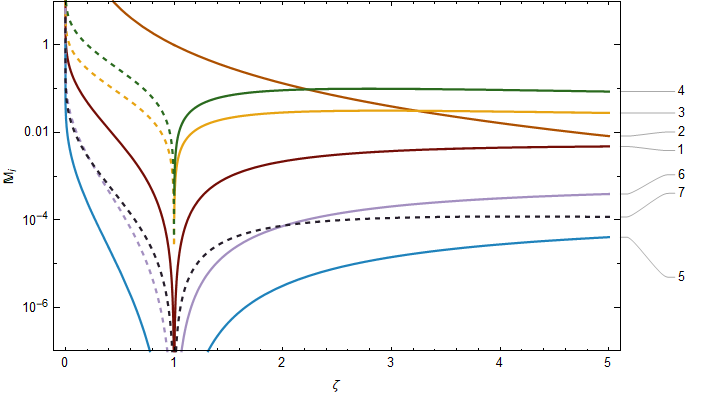}
    \caption{The functions $\mathbb{M}_i(\zeta)$, with solid and dashed lines indicating positive and negative values respectively. Only $\mathbb{M}_2$ is nonzero for $\zeta\to1$.}
    \label{fig:bbMiZeta}
\end{figure}

Much like the $h=1$ case, we may write the mass corrections and charge-to-mass ratio in the canonical ensemble as
\begin{equation}\label{eq:h=1/2bbM}
    \overline{\delta}_iM = -\frac{2}{5q_eq_m}\mathbb{M}_i(\zeta) + \mathcal{O}(T) \,, \qquad z_\text{ext} = 1 + \frac{2}{5q_eq_m}\alpha_i\mathbb{M}_i(\zeta)\,,
\end{equation}
where the functions $\mathbb{M}_i(\zeta)$ are plotted in figure~\ref{fig:bbMiZeta}: the expressions are left to appendix~\ref{app:h=1/2MassCorrections}. Despite the functional form of the $\mathbb{M}_i$ being quite different from those of the $\mathcal{M}_i$, there is a striking similarity between their behavior as $\zeta$ varies. From these mass corrections, we obtain the consistency condition required by the weak gravity conjecture:
\begin{equation}
    \mathcal{C}_\text{dyon}^{h=\frac{1}{2}}(\alpha_i;\zeta) \equiv \alpha_i\mathbb{M}_i(\zeta) > 0 \,,
\end{equation}
where again the $\alpha_1$, $\alpha_2$ and $\alpha_5$ contributions to $\mathcal{C}_\text{dyon}^{h=\frac{1}{2}}$ are always positive and the $\alpha_7$ contribution is always negative. The equal-charge and magnetic limits agree with those found above:
\begin{equation}
    \lim_{\zeta\to1}\mathcal{C}_\text{dyon}^{h=\frac{1}{2}}(\alpha_i;\zeta) = \mathcal{C}_{Q_e=Q_m}(\alpha_i)\,,\qquad \lim_{\zeta\to\infty}(16\zeta)\,\mathcal{C}_\text{dyon}^{h=\frac{1}{2}}(\alpha_i;\zeta) = \mathcal{C}_\text{mag}^{h\neq1}(\alpha_i)\big|_{h=\frac{1}{2}}\,.
\end{equation}
Inverting
\begin{align}
    M &= \frac{1}{\sqrt{2}}\big(Q_e^{2/3}+Q_m^{2/3}\big)^{3/2}\left(1 - \frac{2}{5q_eq_m}\mathcal{C}_\text{dyon}^{h=\frac{1}{2}}(\alpha_i;\zeta) + \frac{1}{2}\widetilde{\mathcal{T}}^2 + \cdots \right) \notag
\end{align}
for $\widetilde{\mathcal{T}}$ when $z=(Q_e^{2/3}+Q_m^{2/3})^{3/2}/\sqrt{2}M=1$ gives
\begin{equation}
    \widetilde{\mathcal{T}} \approx \sqrt{\frac{4\mathcal{C}_\text{dyon}^{h=\frac{1}{2}}(\alpha_i;\zeta)}{5q_eq_m}} \,.
\end{equation}
The entropy correction is then
\begin{equation}
    \Delta S\big|_{z=1} = \frac{4\pi}{\sqrt{5}}\big(\zeta^{1/3}+\zeta^{-1/3}\big)\sqrt{Q_eQ_m\,\mathcal{C}_\text{dyon}^{h=\frac{1}{2}}(\alpha_i;\zeta)} + \mathcal{O}(\alpha_i)\,,
\end{equation}
with the leading term coming from $\Delta S_0$ only.

\subsection{Comments on Entropy Corrections}
In all of the cases considered above we have found entropy corrections to extremal black holes which are not $\mathcal{O}(\alpha_i)$, which we may interpret as arising from the stretching of the black hole horizon. For example, before the introduction of higher-derivative corrections the dyonic, $h=1$ case at extremality has
\begin{equation}
    f(r) = \frac{r^2}{q_eq_m} - \frac{(q_e+q_m)r^3}{q_e^2q_m^2} + \cdots\,,
\end{equation}
i.e.~the degenerate horizon is at $r=0$. $\mathcal{O}(\alpha_i)$ corrections to $f(r)$ lead to $\Delta r\sim\sqrt{\alpha_i}$. In our thermodynamic approach the temperature is tied to the separation between the two horizons via equation~\eqref{eq:EMdTemp}, and so we are still able to capture this square-root behavior in equation~\eqref{eq:h=1entropy} even without computing perturbations to the metric. We may reconcile these observations with equation~\eqref{eq:EntropyAction} by noting that $I_{\text{E},i}(M,Q_e,Q_m)$ often diverges as $(z-1)^{-1/2}$, so that one must go to the next order in $\alpha_i$, $z=1+\mathcal{O}(\alpha_i)$, to achieve a finite answer. The $\mathcal{O}(\alpha_i^{2-h})$ behavior in the magnetic case stems from $f(r)$ not being quadratic near $r=0$ when $h<2$.

\section{The Weak Gravity Conjecture}
\label{sec:WGC}
Having obtained consistency conditions on the Wilson coefficients of the higher-derivative terms for EMd theory, we now turn to the task of showing that these conditions are satisfied under generic assumptions on the UV theory. For this discussion we find it useful to reinstate factors of $M_\text{Pl}$. Before moving to particular examples, it is worth noting that renormalization group effects from graviton, photon and dilaton loops lead to the running of the $\alpha_i$. These Wilson coefficients have dimensions which may be computed perturbatively, with only one or more being most important as one runs to the deep IR. Presently, however, we will show that for a number of generic UV completions of the EMd EFT the weak form of the WGC holds.

\subsection{Unitarity}
Assuming that gravitational effects are subdominant, unitarity requires that both $\alpha_1$ and $\alpha_2$ be non-negative. Here we will show that a similar condition applies also to $\alpha_5$ and $\alpha_7$:
\begin{align}
    \alpha_5,\alpha_7\geq0 \,.
\end{align}
We use the spinning polynomial basis, $P_{\mathbf{s}_n}^{1234}(x)$, of \cite{Arkani-Hamed:2017jhn} and assume that no exchanged particles are massless. Factorization implies the following form for two of the forward helicity amplitudes:
\begin{align}
    \mathcal{M}(\phi\phi\phi\phi) &= \sum_n\Big[\frac{g_{00n}^2}{m_n^2-s}P_{\mathbf{s}_n}^{0000}(1) + \frac{g_{00n}^2}{m_n^2+s}P_{\mathbf{s}_n}^{0000}(1) + a_n+b_ns \Big]\\
    &= \Big(\sum_n\frac{2g_{00n}^2}{m_n^6}\Big)s^2+\cdots \,, \notag\\
    \mathcal{M}(\phi A^\pm\phi A^\mp) &= \sum_n\Big[ \frac{g_{0\pm n}^2}{m_n^2-s}P_{\mathbf{s}_n}^{0\pm0\mp}(1) + \frac{g_{0\pm n}^2}{m_n^2+s}P_{\mathbf{s}_n}^{0\pm0\mp}(1) + a_n+b_ns \Big]\\
    &= \Big( \sum_n\frac{2g_{0\pm n}^2}{m_n^6}\frac{\mathbf{s}_n+1}{\mathbf{s}_n} \Big)s^2 + \cdots \,.\notag
\end{align}
The potentially dangerous contribution from intermediate spin-0 particles is avoided since in that case the coupling $g_{0\pm n}$ is forbidden by locality. On the other hand, the higher-derivative terms of \eqref{eq:hdAction} generate
\begin{equation}
    \mathcal{M}(\phi\phi\phi\phi) \propto \frac{\alpha_5}{M_\text{Pl}^4}s^2 + \cdots \,,\qquad \mathcal{M}(\phi A^\pm\phi A^\mp) \propto \frac{\alpha_7}{M_\text{Pl}^4}s^2 + \cdots \,,
\end{equation}
with positive constants of proportionality. From these we may match the $s^2$ coefficients and read off
\begin{align}
    \alpha_5 \propto \sum_n\frac{g_{00n}^2}{m_n^6} \geq 0\,, \qquad \alpha_7 \propto \sum_n\frac{g_{0\pm n}^2}{m_n^6}\frac{\mathbf{s}_n+1}{\mathbf{s}_n} \geq 0\,.
\end{align}
It is interesting to note that the positivity of $\alpha_7$ implies a contribution from $\alpha_7(\partial\phi\partial\phi FF)$ to the charge-to-mass ratio which is \textit{always} negative. Conceivably this contribution to $z$ could dominate and lead to a violation of the WGC. This shows that in general unitarity of the S-matrix is not sufficient to ensure that the weak form of the WGC holds. This is similar to the situation in \cite{Charles:2019qqt}, where fine-tuning of non-minimal couplings \textit{allows} for the possibility of violating the WGC when running the Wilson coefficients into the deep IR, but this is not borne out in examples.

\subsection{Neutral Scalars}
\label{sec:NeutralScalars}
In the EMd EFT the graviton, photon and dilaton are all massless. In UV completions for which the next-lightest fields are neutral (pseudo)scalars, integrating out these fields at tree-level generates nonzero $\alpha_1$, $\alpha_2$, $\alpha_5$ and $\alpha_6$. Specifically, if one has
\begin{equation}
\begin{aligned}
    \mathcal{L}_\chi &= -\frac{1}{2}(\partial\chi)^2 - \frac{1}{2}m_\chi^2\chi^2 + \frac{\chi}{f_\chi}e^{-3\lambda\phi}\big(F^2\big) + g_\chi M_\text{Pl}\,\chi \,e^{-\lambda\phi}(\partial\phi)^2 \,,\\
    \mathcal{L}_a &= -\frac{1}{2}(\partial a)^2 - \frac{1}{2}m_a^2a^2 + \frac{a}{f_a}e^{-3\lambda\phi}\big(F\widetilde{F}\big)\,,
\end{aligned}
\end{equation}
then one finds
\begin{align}
    \alpha_1 = \frac{2M_\text{Pl}^4}{m_\chi^2f_\chi^2} \,,\qquad \alpha_2=\frac{2M_\text{Pl}^4}{m_a^2f_a^2}\,,\qquad \alpha_5=\frac{2g_\chi^2M_\text{Pl}^2}{m_\chi^2}\,,\qquad \alpha_6 = \frac{4g_\chi M_\text{Pl}^3}{m_\chi^2f_\chi}\,.
\end{align}
With these we find that all of the consistency conditions as required by the WGC are satisfied:
\begin{subequations}
\begin{align}
    \mathcal{C}_\text{mag}^{h\neq1}(\alpha_i) &= \frac{M_\text{Pl}^2}{4m_\chi^2}\left[\frac{4M_\text{Pl}}{f_\chi} + (2-h)g_\chi\right]^2 > 0\,,\\
    \mathcal{C}_\text{el}(\alpha_i;m_\phi) &= \frac{4M_\text{Pl}^4}{m_\chi^2f_\chi^2} + \frac{\lambda^2M_\text{Pl}^2}{m_\phi^2} > 0 \,,\\
    \mathcal{C}_{Q_e=Q_m}(\alpha_i) &= \frac{2M_\text{Pl}^4}{m_a^2f_a^2} > 0 \,,\\
    \mathcal{C}_\text{dyon}^{h=1}(\alpha_i;\zeta) &= \frac{2M_\text{Pl}^4}{m_a^2f_a^2}\mathcal{M}_2(\zeta)\\
    &\quad {}+ \frac{2M_\text{Pl}^4}{m_\chi^2f_\chi^2}\left[\mathcal{M}_1(\zeta) + \left(\frac{g_\chi f_\chi}{M_\text{Pl}}\right)^2\mathcal{M}_5(\zeta) + 2\left(\frac{g_\chi f_\chi}{M_\text{Pl}}\right)\mathcal{M}_6(\zeta)\right] > 0 \,, \notag\\
    \mathcal{C}_\text{dyon}^{h=\frac{1}{2}}(\alpha_i;\zeta) &= \frac{2M_\text{Pl}^4}{m_a^2f_a^2}\mathbb{M}_2(\zeta)\\
    &\quad {}+ \frac{2M_\text{Pl}^4}{m_\chi^2f_\chi^2}\left[\mathbb{M}_1(\zeta) + \left(\frac{g_\chi f_\chi}{M_\text{Pl}}\right)^2\mathbb{M}_5(\zeta) + 2\left(\frac{g_\chi f_\chi}{M_\text{Pl}}\right)\mathbb{M}_6(\zeta)\right] > 0 \,. \notag
\end{align}
\end{subequations}
These rely on the following facts,
\begin{equation}
\begin{aligned}
    \mathcal{M}_2(\zeta) &> 0 \qquad \forall\zeta\,,\\
    \mathbb{M}_2(\zeta) &> 0 \qquad \forall\zeta\,,\\
    \mathcal{M}_1(\zeta) + x^2\mathcal{M}_5(\zeta) + 2x\mathcal{M}_6(\zeta) &\geq 0 \qquad \forall x,\zeta\,,\\
    \mathbb{M}_1(\zeta) + x^2\mathbb{M}_5(\zeta) + 2x\mathbb{M}_6(\zeta) &\geq 0 \qquad \forall x,\zeta\,,
\end{aligned}
\end{equation}
the last two of which are nontrivial: $x$, $\mathcal{M}_6(\zeta)$ and $\mathbb{M}_6(\zeta)$ may be of either sign.

\subsection{Charged Scalars and Fermions}
\label{sec:Charged}
\begin{figure}[t]
    \centering
    \begin{subfigure}{0.23\textwidth}
    \centering
    \begin{fmffile}{diagram1}
    \begin{fmfgraph}(90,80)
        \fmfleft{i1,i2}
        \fmfright{o1,o2}
        \fmf{fermion,tension=0.7}{v1,v2,v3,v4,v1}
        \fmf{photon}{i1,v1} \fmf{photon}{i2,v2}
        \fmf{photon}{o1,v4}
        \fmf{photon}{o2,v3}
    \end{fmfgraph}
    \end{fmffile}
    \end{subfigure}
    \begin{subfigure}{0.23\textwidth}
    \centering
    \begin{fmffile}{diagram2}
    \begin{fmfgraph}(90,80)
        \fmfleft{i1,i2}
        \fmfright{o1,o2}
        \fmf{fermion,tension=0.4}{v1,v2,v3,v1}
        \fmf{dbl_wiggly}{v3,v4}
        \fmf{photon}{i1,v1} \fmf{photon}{i2,v2}
        \fmf{photon}{o1,v4}
        \fmf{photon}{o2,v4}
    \end{fmfgraph}
    \end{fmffile}
    \end{subfigure}
    \begin{subfigure}{0.23\textwidth}
    \centering
    \begin{fmffile}{diagram3}
    \begin{fmfgraph}(90,80)
        \fmfleft{i1,i2}
        \fmfright{o1,o2}
        \fmf{fermion,tension=0.7}{v1,v2,v3,v4,v1}
        \fmf{photon}{i1,v1} \fmf{photon}{i2,v2}
        \fmf{dashes}{o1,v4}
        \fmf{dashes}{o2,v3}
    \end{fmfgraph}
    \end{fmffile}
    \end{subfigure}
    \begin{subfigure}{0.23\textwidth}
    \centering
    \begin{fmffile}{diagram4}
    \begin{fmfgraph}(90,80)
        \fmfleft{i1,i2}
        \fmfright{o1,o2}
        \fmf{fermion,tension=0.7}{v1,v2,v3,v4,v1}
        \fmf{dashes}{i1,v1} \fmf{dashes}{i2,v2}
        \fmf{dashes}{o1,v4}
        \fmf{dashes}{o2,v3}
    \end{fmfgraph}
    \end{fmffile}
    \end{subfigure}
    \caption{Examples of 1-loop diagrams leading to higher-derivative interactions.}
    \label{fig:1loop}
\end{figure}
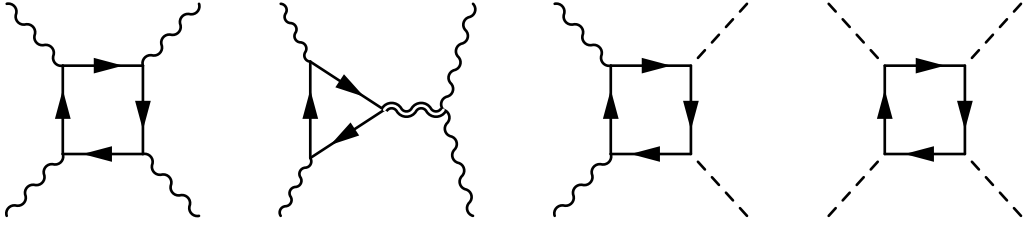
If the lowest-lying states in the UV theory are charged scalars and/or fermions, then the leading contributions to the $\alpha_i$ are generated by one-loop diagrams, such as those in figure~\ref{fig:1loop}. Assuming weak coupling at the scale of these charged states, we need consider only electrically-charged particles. Writing $z\sim\frac{q_e}{m}$ for the charge-to-mass ratio of such a particle, we may estimate
\begin{equation}
\begin{aligned}
    |\alpha_{1,2}| &\sim \max\{\mathcal{O}(1),\mathcal{O}(z^2),\mathcal{O}(z^4)\} \,,\\
    |\alpha_{3,6,7}| &\sim \max\{\mathcal{O}(1),\mathcal{O}(z^2)\} \,,\\
    |\alpha_{4,5}| &\sim \mathcal{O}(1) \,.
\end{aligned}
\end{equation}
For $z\gg1$, electromagnetic effects dominate and unitarity ensures that $\alpha_1$ and $\alpha_2$ are positive. Of course, if $z\gg1$ then the WGC is already satisfied without considering black hole states, but in this limit we do find that the consistency conditions are satisfied:
\begin{subequations}\label{eq:ChargedConditions}
\begin{align}
    \mathcal{C}_\text{mag}^{h\neq1}(\alpha_i) &\approx 2\alpha_1 > 0 \,,\\
    \mathcal{C}_\text{el}(\alpha_i;m_\phi) &\approx 2\alpha_1 + \frac{\lambda^2}{m_\phi^2} > 0 \,,\\
    \mathcal{C}_{Q_e=Q_m}(\alpha_i) &= \alpha_2 > 0 \,,\\
    \mathcal{C}_\text{dyon}^{h=1}(\alpha_i;\zeta) &\approx \alpha_1\mathcal{M}_1(\zeta) + \alpha_2\mathcal{M}_2(\zeta) > 0\,,\\
    \mathcal{C}_\text{dyon}^{h=\frac{1}{2}}(\alpha_i;\zeta) &\approx \alpha_1\mathbb{M}_1(\zeta) + \alpha_2\mathbb{M}_2(\zeta) > 0 \,,
\end{align}
\end{subequations}
where we have used that $\mathcal{M}_1$, $\mathcal{M}_2$, $\mathbb{M}_1$ and $\mathbb{M}_2$ are positive for all $\zeta$. The above conditions ensure that these black hole states are unstable and can decay to smaller dyonic black holes, as was the original motivation for the WGC.

\subsection{Open String-like UV Completion}
\label{sec:Regge}
Suppose now that the UV theory has no low-lying states, but rather towers of higher-spin, Regge states accompanying the graviton, photon and dilaton. Writing $\Lambda_\text{QFT}<M_\text{Pl}$ for the scale at which quantum field theory breaks down, one would expect the following hierarchy:
\begin{equation}
    |\alpha_{1,2,5,6,7}|\sim\mathcal{O}\bigg(\frac{M_\text{Pl}^4}{\Lambda_\text{QFT}^4}\bigg) \,,\qquad |\alpha_3|\sim\mathcal{O}\bigg(\frac{M_\text{Pl}^2}{\Lambda_\text{QFT}^2}\bigg) \,,\qquad |\alpha_4|\sim\mathcal{O}(1)\,.
\end{equation}
By itself such a hierarchy is not enough to guarantee that the WGC conditions are satisfied, even if supplemented with $\alpha_1,\alpha_2,\alpha_5,\alpha_7\geq0$. While the contributions to the charge-to-mass ratio from the $\alpha_1$, $\alpha_2$ and $\alpha_5$ terms are always positive, the $\alpha_7$ contribution is always negative and the $\alpha_6$ contribution changes sign with $\zeta$ larger or smaller than one. It is enough, however, to have
\begin{equation}
    \max\{\alpha_1,\alpha_2,\alpha_5\}\gtrsim |\alpha_6|,\alpha_7 \geq 0 \,.
\end{equation}
Such an inequality \textit{is} found in open string-like UV completions, where the Regge states of the photon are open string states, while the Regge states of the graviton and dilaton are closed string states. Since $g_\text{s}\sim g_\text{open}^2$, the contributions from each sector to the Wilson coefficients are then
\begin{equation}
    [\alpha_{1,2}]_\text{open} \sim \frac{M_\text{Pl}^2}{g_\text{s}M_\text{s}^2} \,, \qquad [\alpha_{1,2}]_\text{closed} \sim [\alpha_{3,4,5,6,7}]_\text{open}\sim [\alpha_{3,4,5,6,7}]_\text{closed} \sim \frac{M_\text{Pl}^2}{M_\text{s}^2} \,.
\end{equation}
Given $g_\text{s}\ll1$, $\alpha_1$ and $\alpha_2$ dominate and the WGC conditions are satisfied, just as in~\eqref{eq:ChargedConditions}.

\subsection{The Heterotic String}
Here we quickly check that the Wilson coefficients derived from compactifying heterotic string theory down to 4D satisfy the conditions found for $h=1$. In string frame the $\mathcal{O}(\alpha')$ heterotic string action reads \cite{Gross:1986mw}
\begin{equation}
\begin{aligned}
	I_{10} &= \frac{M_{10}^8}{2}\int\d[10]{x}\,\sqrt{-G}\,e^{-2\Phi}\Big[ R + 4(\partial\Phi)^2 - \frac{1}{2}F_{MN}F^{MN}\\
	&\qquad + \frac{\alpha'}{8}\Big( R^{MNLP}R_{MNLP} + \frac{3}{4}(F_{MN}F^{MN})^2 + \frac{3}{4}(F_{MN}\widetilde{F}^{MN})^2 \Big) \Big]\,.
\end{aligned}
\end{equation}
Dimensionally reducing to 4D on $\mathcal{M}_{10}=\mathcal{M}_4\times X_6$ and rescaling the dilaton to $\phi=\sqrt{2}\Phi$ leads to
\begin{equation}
\begin{aligned}
	I_4 &= \int\d[4]{x}\,\sqrt{-g}\,e^{-\sqrt{2}\phi}\Big[ \frac{1}{2}R + (\partial\phi)^2 - \frac{1}{4}(F^2)\\
	&\qquad\qquad + \frac{\alpha'}{16}\Big( R^{\mu\nu\rho\sigma}R_{\mu\nu\rho\sigma} + \frac{3}{4}(F^2)^2 + \frac{3}{4}(F\widetilde{F})^2 \Big) \Big] \,,
\end{aligned}
\end{equation}
where $M_\text{Pl}^2=M_{10}^8\operatorname{vol}(X_6)=1$. In Einstein frame the above becomes
\begin{equation}
\begin{aligned}
	I_4 &= \int\d[4]{x}\,\sqrt{-g}\Big[ \frac{1}{2}R - \frac{1}{2}(\partial\phi)^2-\frac{1}{4}e^{-\sqrt{2}\phi}(F^2) + \frac{\alpha'}{16}\Big( \frac{3}{2}e^{-3\sqrt{2}\phi}(F^2)^2\\
	&\qquad\qquad +\frac{7}{4}e^{-3\sqrt{2}\phi}(F\widetilde{F})^2+e^{-\sqrt{2}\phi}R_\text{GB} + 2e^{-2\sqrt{2}\phi}(F^2)(\partial\phi)^2 \Big) \Big]\,.
\end{aligned}
\end{equation}
In particular,
\begin{align}
	(\alpha_1,\alpha_2,\alpha_3,\alpha_4,\alpha_5,\alpha_6,\alpha_7) = \frac{\alpha'}{16}\big(6,\,7,\,0,\,2,\,0,\,8,\,0\big) \,,
\end{align}
which indeed ensure $\mathcal{C}_\text{dyon}^{h=1}(\alpha_i;\zeta)>0$ for all $\zeta$.\footnote{A similar story for the dimensional reduction of 5D Gauss-Bonnet gravity, $\frac{1}{2}\int\d[5]{x}\,\sqrt{-g}\big(R + \alpha R_\text{GB}\big)$, leads to $\alpha_i = \frac{\alpha}{12}(66,27,-12,12,32,74,96)$ and $\mathcal{C}_\text{dyon}^{h=\frac{1}{2}}(\alpha_i;\zeta)>0$ for all $\zeta$ when $\alpha>0$.}

\section{Discussion}
\label{sec:Discussion}
In this paper we have calculated higher-derivative corrections to Einstein-Maxwell-dilaton black holes for a variety of choices for electric charge, magnetic charge and dilaton coupling constant. Motivated by the swampland program and the weak gravity conjecture in particular, we found constraints on the Wilson coefficients of the effective theory which ensure that the charge-to-mass ratio of black holes increases from its classical value. For electrically charged black holes perturbative control is lost due to the classically vanishing area and diverging dilaton at extremality.

By considering several generic UV completions of EMd theory we have shown that the consistency conditions imposed by the WGC are generically satisfied. These checks show that the charge-to-mass ratio of extremal, dilatonic black holes increases from its classical value for a range of electric, magnetic and dyonic solutions. We have focused on those cases where we can obtain closed-form expressions, but much of the parameter space remains unchecked. For dyonic black holes and general coupling $\lambda$, one could check numerically that similar results hold. Given the similarity of the $h=1$ and $h=1/2$ cases, we expect that nothing drastically different would be found for general $h$.

Our work provides more nontrivial evidence for the WGC as a general constraint for identifying quantum gravity-derived EFTs. Even in this more general setting all large black holes are unstable to decay, either through thermal radiation if at finite temperature or through the kinematically allowed emission of a superextremal black hole. For the heterotic string, the weak form of the WGC pursued here is connected via modular invariance to a strong form where the superextremal states are light \cite{Aalsma:2019ryi}.

Refs.~\cite{Charles:2019qqt} and \cite{Jones:2019nev} have recently shown that one-loop contributions to the Wilson coefficients generically lead to the weak form of the WGC being satisfied in the deep IR. It would be interesting to investigate this argument with the exponential coupling of the dilaton considered here. In addition, our setup can also be extended to a supersymmetric one by including an axion in addition to the dilaton. In the absence of non-perturbative effects the $\operatorname{SL}(2,\mathbb{R})$ symmetry leads to \textit{all} dyonic solutions having vanishing classical area, as we had here in the pure electric and magnetic cases. We leave such considerations for future work.

We conclude with a comment on the difficulty of demonstrating the WGC using positivity bounds for scattering amplitudes. The main obstruction to deriving the mild WGC from the positivity bounds is the $t$-channel graviton exchange $\sim s^2/t$. In~\cite{Hamada:2018dde} it was clarified under which conditions the positivity of the $\mathcal{O}(s^2)$ coefficient is justified and thus the mild WGC follows by carefully studying contributions from Regge states. More recently, Ref.~\cite{Bellazzini:2019xts} proposed a regularization scheme based on compactification of 4D gravitational theory to 3D. Even though it was claimed that it leads to the $\mathcal{O}(s^2)$ positivity and thus the mild WGC in general setups, several big assumptions are in order: First of all, to remove the $t$-channel singularity in their scenario, one needs to take the forward limit $t\to0$ first and then take the decompactification limit, which is an opposite ordering to obtaining the 4D bound. Second, this scenario is motivated by a potential non-perturbative UV completion of 3D gravity~\cite{Ciafaloni:1992hu,tHooft:1988qqn,Deser:1993wt,Zeni:1993ew}. However, it is far from obvious if the same scenario works in the 3D gravitational theory with a 4D origin. For example, if we assume a perturbative UV completion of gravity just as sting theory, the KK reduced 3D theory will contain infinitely many higher-spin Regge states. As demonstrated in~\cite{Hamada:2018dde}, the standard derivation of the positivity bounds~\cite{Adams:2006sv} cannot be justified unless effects of the Regge states are subdominant. Therefore, more studies on Regge states will be encouraged to understand how to resolve the $t$-channel singularity and complete the proof of the mild WGC, at least as long as we consider the string theory type UV completion of gravity.

\acknowledgments

TN would like to thank the University of Wisconsin for their hospitality, where a part of this work was done.
The work of GL and GS is supported in part by the DOE grant de-sc0017647 and the Kellett Award of the University of Wisconsin.
TN is supported in part by JSPS KAKENHI Grant Numbers JP17H02894 and JP18K13539, and MEXT KAKENHI Grant Number JP18H04352.

\appendix
\section{\texorpdfstring{Dyonic, $\mathbf{h=1/2}$ Mass Corrections}{h=1/2 Mass Corrections}}
\label{app:h=1/2MassCorrections}
For $\lambda^2=3/2$ ($h=1/2$) the mass corrections in the canonical ensemble are (see equation~\eqref{eq:h=1/2bbM})
\begin{equation}
    \overline{\delta}_iM = -\frac{2}{5q_eq_m}\mathbb{M}_i(\zeta) + \mathcal{O}(T) \,.
\end{equation}
The functions $\mathbb{M}_i(\zeta)$ are, introducing $x\equiv\zeta^{2/3}$,
\begin{subequations}
\begin{align}
    \mathbb{M}_1(\zeta) &= \frac{1}{8\sqrt{x}(1-x)^2(1+x)}\bigg[(96-52x+x^2)\\
    &\hspace{138pt} - 15(4+x)\frac{\operatorname{arccosh}{x}}{\sqrt{x^2-1}} + 30x^2\,\frac{\operatorname{arcsech}{x}}{\sqrt{1-x^2}}\bigg] \,, \notag\\[1em]
    \mathbb{M}_2(\zeta) &= \frac{1}{2x^{7/2}(1-x)^2(1+x)}\bigg[(6+8x+x^2)-15x\,\frac{\operatorname{arccosh}{x}}{\sqrt{x^2-1}}\bigg] \,,\\[1em]
    \mathbb{M}_3(\zeta) &= \frac{1}{48\sqrt{x}(1-x)^2(1+x)}\bigg[(154-208x+9x^2)- 120(1-x)\log{x}\\
    &\hspace{52pt} - 15(4-7x+4x^2)\frac{\operatorname{arccosh}{x}}{\sqrt{x^2-1}}-60 (2-2x-x^2)\frac{\operatorname{arcsech}{x}}{\sqrt{1-x^2}} \bigg] \,, \notag\\[1em]
    \mathbb{M}_4(\zeta) &= \frac{3}{64\sqrt{x}(1-x)^2(1+x)}\bigg[ (224-128x+9x^2) - 40(5-4x)\log{x}\\
    &\hspace{30pt} - 5(16-17x+16x^2)\frac{\operatorname{arccosh}{x}}{\sqrt{x^2-1}} - 10(20-16x-x^2)\frac{\operatorname{arcsech}{x}}{\sqrt{1-x^2}}\bigg] \,, \notag\\[1em]
    \mathbb{M}_5(\zeta) &= -\frac{9}{512\sqrt{x}(1-x)^2(1+x)}\bigg[ (184-3x-x^2) + 40(1-2x)\log{x}\\
    &\hspace{52pt} -45(3+x-2x^2)\frac{\operatorname{arccosh}{x}}{\sqrt{x^2-1}} +10(4-8x-5x^2)\frac{\operatorname{arcsech}{x}}{\sqrt{1-x^2}}\bigg] \,, \notag\\[1em]
    \mathbb{M}_6(\zeta) &= \frac{1}{128\sqrt{x}(1-x)^2(1+x)}\bigg[ (136+353x+6x^2) + 240(1-x)\log{x}\\
    &\hspace{30pt} - 15(13+14x-14x^2)\frac{\operatorname{arccosh}{x}}{\sqrt{x^2-1}} + 60(4-4x-5x^2)\frac{\operatorname{arcsech}{x}}{\sqrt{1-x^2}}\bigg] \,, \notag\\[1em]
    \mathbb{M}_7(\zeta) &= \frac{15}{64\sqrt{x}(1-x)^2(1+x)}\bigg[ 5(1+x)+4(1-x)\log{x}\\
    &\hspace{69pt} - (5+4x-4x^2)\frac{\operatorname{arccosh}{x}}{\sqrt{x^2-1}} +(4-4x-5x^2)\frac{\operatorname{arcsech}{x}}{\sqrt{1-x^2}} \bigg] \,. \notag
\end{align}
\end{subequations}
Principal values should be used for the square-root and inverse hyperbolic functions: branch cuts conspire to make $\operatorname{arccosh}{x}/\sqrt{x^2-1}$ and $\operatorname{arcsech}{x}/\sqrt{1-x^2}$ smooth, real functions of $x>0$. The functions $\mathbb{M}_i(\zeta)$ are plotted in figure~\ref{fig:bbMiZeta}.

\end{document}